# Spin-orbit interaction driven terahertz nonlinear dynamics in transition metals

Ruslan Salikhov[1]*, Markus Lysne[2]*, Philipp Werner[2], Igor Ilyakov[1], Michael Schüler[2,3], Thales V. A. G. de Oliveira[1], Alexey Ponomaryov[1], Atiqa Arshad[1], Gulloo Lal Prajapati[1], Jan-Christoph Deinert[1], Pavlo Makushko[1], Denys Makarov[1], Thomas Cowan[1], Jürgen Fassbender[1,4], Jürgen Lindner[1], Aleksandra Lindner[1], Carmine Ortix[5], and Sergey Kovalev[1,6]*

[1]Helmholtz-Zentrum Dresden-Rossendorf, Dresden, Germany

[2]Department of Physics, University of Fribourg, Fribourg, Switzerland

[3] Laboratory for Materials Simulations, Paul Scherrer Institute, Villigen, Switzerland

[4]Technische Universität Dresden, Dresden, Germany

[5]Dipartimento di Fisica "E.R. Caianiello", Universita di Salerno, Salerno, Italy

[6] Fakultät Physik, Technische Universität Dortmund, Dortmund, Germany

*Corresponding authors. Email: r.salikhov@hzdr.de, markus.lysne@unifr.ch, sergey.kovalev@tu-dortmund.de

**Abstract:** The interplay of electric charge, spin, and orbital polarizations, coherently driven by picosecond long oscillations of light fields in spin-orbit coupled systems, is the foundation of emerging terahertz spintronics and orbitronics. The essential rules for how terahertz light interacts with these systems in a nonlinear way are still not understood. In this work, we demonstrate a universally applicable electronic nonlinearity originating from spin-orbit interactions in conducting materials, wherein the interplay of light-induced spin and orbital textures manifests. We utilized terahertz harmonic generation spectroscopy to investigate the nonlinear dynamics over picosecond timescales in various transition metal films. We found that the terahertz harmonic generation efficiency scales with the spin Hall conductivity in the studied films, while the phase takes two possible values (shifted by $\pi$), depending on the $d$-shell filling. These findings elucidate the fundamental mechanisms governing non-equilibrium spin and orbital polarization dynamics at terahertz frequencies, which is relevant for potential applications of terahertz spin- and orbital-based devices.

The interaction between terahertz (THz) light and matter has gained significant attention due to its potential for the coherent generation and nonlinear manipulation of charge and spin currents on ultrafast timescales. Notable examples include THz-driven Dirac currents in topological surfaces[1], efficient THz frequency conversion[2,3], ultrafast magnetization switching[4], and short-wavelength magnon excitation via spin-orbit torques[5]. Extensive research is being conducted in the areas of nonlinear THz-frequency dynamics in superconductors[6,7], Dirac materials[8,9], and magnetic heterostructures[10-12]. The THz field-driven nonlinear processes can arise from various phenomena, such as electronic phase transitions that lead to enormous conductivity changes in superconductors[6,7], the non-parabolic band structure of carriers in Dirac materials[3,8,9], and the low heat capacity in graphene necessary for thermodynamic harmonic generation[2]. There is intense research focused on utilizing ultrafast spin-to-charge interconversion, which exploits spin-pumping effects at magnetic/nonmagnetic interfaces as a means of THz harmonic generation[10,11]. Despite significant progress in the research on nonlinear THz-driven dynamics in complex systems, a comprehensive understanding of the interaction between THz light and metals has yet to be established.

It is believed that metallic films exhibit negligible THz nonlinearities due to parabolic electronic band structure, and large electronic heat capacity. The free electron dynamics in metals shows a linear[13] or thermal[14] response at THz frequencies. This is particularly true, when the electronic spin and orbital degrees of freedom are not considered. However, spin-based non-equilibrium processes in transition metals (TMs) are central to the operation of spintronics devices[15,16]. The charge-to-spin conversion via the spin Hall effect and the reciprocal spin-to-charge conversion via the inverse spin Hall effect in TMs remain efficient even at picosecond time scales[5,17-20]. Additionally, a recent study reported ultrafast orbital angular momentum currents in tungsten films, which are converted to the charge current via the inverse orbital Rashba-Edelstein effect[21].

In this work, we reveal a nonlinear THz response in TM films, which originates from ultrafast electron spin dynamics coupled to the non-equilibrium orbital polarization. The coupling results in terahertz third harmonic generation (THG), as demonstrated in various $3d$, $4d$, and $5d$ TMs. We show that the THG amplitude closely resembles the spin Hall conductivity (SHC) values of the studied films, and the polarity of THG follows the sign of the SHC. Our research provides new insights into the general and complex interplay between the nonlinear dynamics of charge carriers, their spin and orbital momenta, which is important for understanding non-equilibrium manipulation of matter.

To investigate the nonlinear carrier dynamics in transition metal films, we utilized THz time-domain spectroscopy (TDS). The schematic of our experiment is illustrated in Fig. 1a. We employed a narrowband THz radiation (referred to as fundamental beam) with a central frequency of 0.5 THz, a bandwidth of about 20%, and a pulse energy of 1.5 µJ (details of the fundamental beam source are provided in the Methods). The radiation was focused on a sample, which consisted of a 4 nm-thin (unless otherwise specified) TM film (nonmagnetic - Pt, Pd, Ir, W, Ta, Nb, Au and magnetic - Co, Py = $Ni_{81}Fe_{19}$), grown on a 1 mm thick quartz glass substrate using magnetron sputter deposition. All samples were capped by a 10-nm-thick $SiO_2$ layer, unless otherwise specified. The generated (and transmitted through the sample) THz radiation at 1.5 THz was refocused on a 2-mm thick ZnTe crystal, and its dynamics was measured using electro-optic sampling. Two bandpass filters with a central frequency of 1.5 THz were placed between the sample and ZnTe crystal to suppress the fundamental radiation.

Figure 1b illustrates the temporal evolution of the fundamental radiation used to excite the metallic samples. The estimated peak field is about 200 kV/cm. Figure 1c shows the measured THz THG in Pd and Au samples, and their corresponding Fourier spectra are presented in Fig. 1e. The Pd film acts as a source for THz THG, while no harmonics are observed in the Au sample. The efficiency of THG in Pd at room temperature is about two orders of magnitude smaller in amplitude when compared to that of topological insulators[3]. The THG signal is linearly polarized, with the polarization collinear to the fundamental beam polarization. The THz THG vanishes for circular polarization of the pump pulse (Supplementary Information Ext. Fig.1).

The intensity of the THz THG is presented for all samples in Fig. 1d. To account for variations in transmission when comparing the samples, the THG intensity was normalized by the cube of the sample's transmission. The THG intensity is highest for the platinum sample, lowest for the niobium sample, while the gold film sample shows no THG signal. The THG signal in the gold film is determined by the measurement noise floor and is similar to that of the bare substrate, representing the sensitivity limit of our experimental setup. We found no impact of magnetic fields with amplitude up to 1 T and direction within the film plane on the THG signal in all samples. The THG of magnetic films such as Py and Co was found to be independent on their magnetization direction (Supplementary Information Ext. Fig.2).

The distribution of the THG intensity across different materials exhibits a qualitative correlation with their SHC characteristics[22]. Similar to the SHC, the THG intensity (Fig. 1d) is found to be the highest for Pt, Pd and Py (Py = $Ni_{81}Fe_{19}$), which belong to the same group-10 elements of the Periodic Table. When moving from the 3$d$ (Py) to the 4$d$ (Pd) and 5$d$ (Pt) elements, the THG intensity increases. This trend is in line with the SHC, which is consistent with the known increase in spin-orbit coupling strength with atomic number[15,22,23]. As one moves through the series of 5$d$ elements (Ta, W, Ir, Pt, Au), the THG intensity distribution in Fig. 1d closely resembles the SHC calculations[22-24]. When considering ferromagnetic metals (Py and Co), it is predicted that Py has a larger SHC compared to Co[22], and this trend is also evident in our diagram in Fig. 1d.

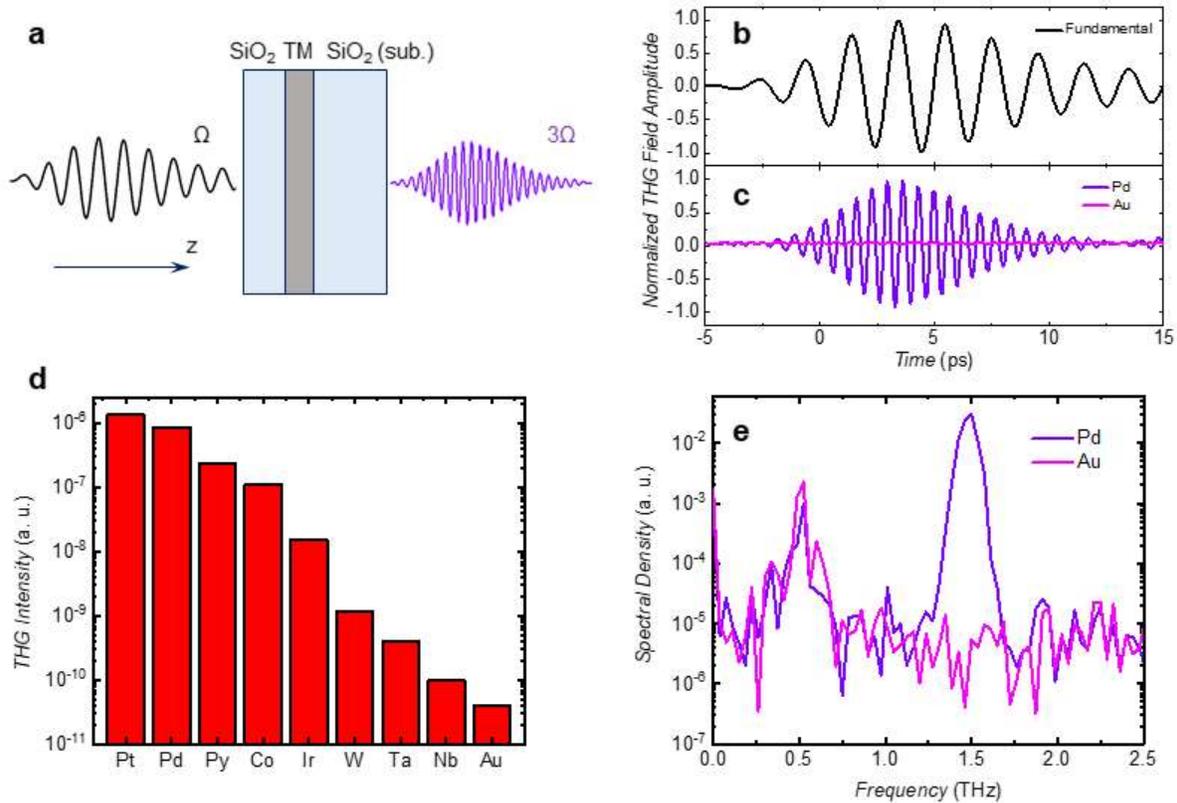

**Fig. 1. Third harmonic generation in transition metals. a**, Schematics of the experiment. Multi-cycle terahertz pulses with a central frequency of Ω (black) are incident on TM films capped with $SiO_2$. The emission of terahertz radiation at the third harmonic frequency, 3Ω (violet), is detected using electro-optical sampling. **b**, Time scan of the incident (fundamental) radiation, measured directly from the source. **c**, Comparison of the THG signals between 4-nm-thin Pd and Au layers. The THG signal from the Pd layer is evident, while the Au layer in our samples exhibits virtually zero THG amplitude. The fast Fourier transform spectra of the corresponding signals are shown in (**e**). The leaking signal from the fundamental beam is visible at 0.5 THz for both samples. Only the Pd layer shows an intense signal at the third harmonic of 1.5 THz. **d**, The comparison of the THG signal intensity, normalized to the cube of the sample's transmission, for various transition-metal films. To ensure consistency, the transmitted fundamental and THG signals from all films were measured under the same experimental conditions. Please, note the logarithmic scale.

Next, we illustrate that the THG phase depends on the *d*-band filling and, consequently, on the sign of the SHC. To demonstrate this relationship, three sets of experiments were conducted using different samples. The results are shown in Fig. 2. Figure 2a compares time scans of the THG signal in the samples. There is a phase shift of about 180 degrees between the Pd and Py group of samples (solid lines) and the Ta and W group (dashed lines). The THG signals of the Pt, Ir, Co, Py (Nb, Ta) films are in phase with that of Pd (W). Note that Nb, Ta, and W have less than half-filled *d*-shells and exhibit a negative SHC. Next, we compared the THG signals from the Co and W samples with a 25-nm thick Bi film, which exhibits strongly spin-orbit coupled surface states[25]. As the THG amplitude in 25-nm thick Bi is 50 times higher compared to 4 nm thin Co and reaches zero below 15 nm film thickness[25], the THG in the Bi film cannot be due to the spin Hall effect, but rather originates from surface carriers with a linear coupling between crystalline momentum and electron spin[26]. In fact, as shown in Fig. 2b, the signals from the Co and W films are shifted by ± 90° with respect to the Bi signal, depending on the sign of their SHC. To eliminate the effect of the crystal structure, we compared the THz THG phase in the body-centered cubic (bcc) Fe film and in the face-centered cubic (fcc) Py film, both grown using molecular beam epitaxy (see Methods for details). Both samples were covered with an Au layer to protect them from oxidation. We found no visible changes in the THG phase within the experimental accuracy, indicating that the THG phase is independent of the crystal structure of the TM (Fig. 2c).

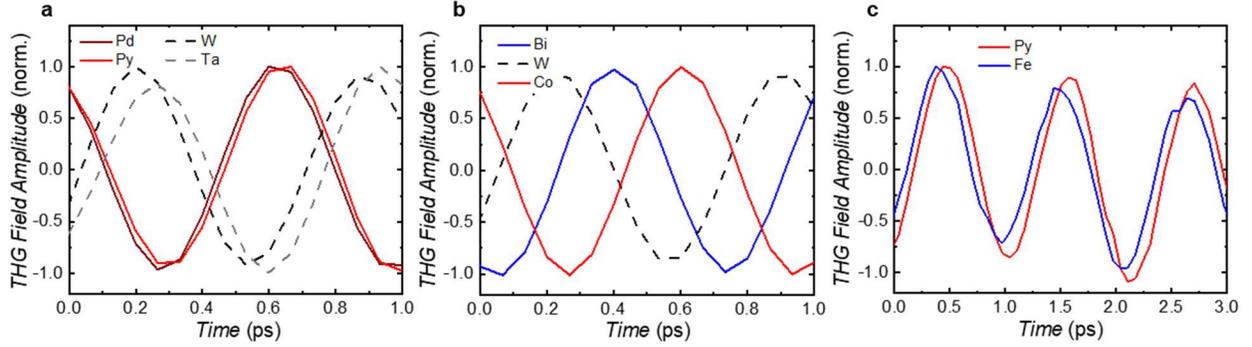

**Fig. 2. Third harmonic radiation phase.** Comparison of THG time scans for (**a**) different transition-metal films; (**b**) W, Co and Bi films; and (**c**) 4-nm-thin Py and Fe films grown on MgO substrates using molecular beam epitaxy and capped with a 3-nm thin Au layer. The films in (**c**) were excited using a 0.3 THz fundamental beam. TMs with negative SHC or less than half-filled *d*-bands are indicated with a dashed line.

We studied THz THG dependence on the fundamental beam fluence, film thickness, and temperature. As shown in Fig. 3a, the THG intensity follows the expected proportional relationship to the cube of the fundamental power, with no saturation even at incident field of 200 kV/cm. To examine the thickness dependence, we used Pd and W samples of varying thicknesses, each capped with a $SiO_2$ layer (Fig. 3b). We normalized the THG intensity to the cube of the sample transmission to account for screening effects (the data without normalization are given in the Supplementary Information, Ext. Fig. 3). The THG field amplitudes in both samples exhibit a linear behavior with thickness. This behavior suggests that the THG processes originate from bulk effects, with no obvious influence of surface contributions. At the Pd film thickness of 2 and 12 nm, no visible THG was detected.

The temperature dependence of THz THG in 4 nm thick films of Pd, Co, Nb, Ta and W was studied using a gas flow cryostat. Due to the limitations of the cryostat's numerical aperture, the field strength of the pump pulse was reduced, resulting in only the Pd and Co films showing THG above the noise floor at room temperature. Nevertheless, all TM films demonstrated a strong increase in THz THG as temperature decreased (Fig. 3c). The THG in the Pd sample was over 60 times higher at 5 K than at 300 K. The phase of THG was found to be temperature-independent.

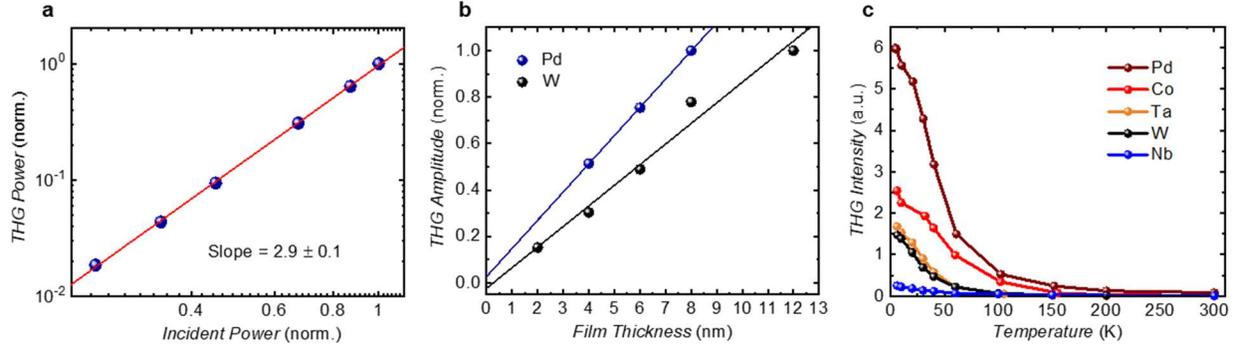

**Fig. 3. Impact of incident power, layer thickness and temperature. a**, Incident power dependence of the third harmonic signal in a Pd film. The red solid line shows the linear fit to the data in the log-log plot. **b**, The effect of Pd and W film thickness on the third harmonic generation amplitude (normalized), after the correction of the screening effect. Solid lines represent linear fits to the data points. **c**, Temperature dependence of terahertz harmonic generation in 4-nm-thick transition metal films.

**Discussion**

The intensity and phase of the harmonic generation in different TMs resemble their spin Hall characteristics, indicating that spin-orbital effects determine the nonlinear interactions between THz light and metallic films. The THz THG is independent on both the applied magnetic field (up to 1 T), magnetization direction (for Co, Fe and Py) and crystal structure (bcc Fe versus fcc Py). The linear thickness dependence demonstrates that THG is a bulk effect. We show in the Supplementary Information that THG from surface states is not expected to generally display a $\pi$ phase shift. The filling of the $d$ bands is reflected in a change of the Rashba coupling characterizing the surface states. This does neither change the carrier velocities nor the geometrical properties of the electronic wave-functions affecting nonlinear electromagnetic properties, including the Berry connection polarizability entering the third-order current response.

Our numerical simulations shown in Supplementary Information, based on relativistic density functional theory bandstructures, indicate that the intrinsic electric or spin current nonlinearities cannot explain a $d$-band filling dependent 180° phase shift of THz THG. This suggests that the THz THG in TMs cannot be solely described by the nonlinear interaction between electrical and spin currents, and that an additional dynamical THz-induced effect needs to be considered. This effect must be odd in the electron momentum (or electrical current direction), since an even effect combined with electron spin would result in second harmonic generation[27,28] or optical rectification[29]. A possible candidate could at first glance be the THz field induced spin Nernst effect[22,30,31]. However, the spin Nernst conductivity (SNC) in Pt and W have opposite signs compared to their SHC[22,30,31], while in Pd and Ta the SNC has the same sign as the SHC[22].

We derive a phenomenological wave equation for THG that accounts for both the SHC and its sign (see Supplementary Information). The THz electric field (**E**) generates electrical currents in the TM films (Fig. 4), oscillating with the frequency of the THz fundamental pulse. Due to the spin Hall effect, the electrical currents induce a spin angular momentum ($\vec{s}$) polarized perpendicular to **E**. The spin direction circulates with either right or left handedness, depending on the sign of the spin Hall angle (black arrow in Fig. 4). We introduce an additional THz field induced parameter, X, which has a similar symmetry to the SHE-induced dynamical spin texture (blue arrow in Fig. 4). The interaction between the electron spins and X is a source of the odd nonlinear response. Since the THz THG changes polarity depending on the sign of the SHC, we assume that the Hamiltonian of the interaction involves a scalar product of $\vec{s}$ and $\vec{X}$, $H_X \propto \int (\vec{s}, \vec{X}) dV$ (where V is the volume of integration), and that the X-vector is independent of the sign of the SHC. The Hamiltonian $H_X$ dynamically modify the samples' conductivity, leading to the THz THG.

Considering all characteristics required for the THz THG, the best candidate is the orbital Hall effect (OHE), as has been proven recently for Ti films[36]. Similar to the spin current $J_S = \sigma_{SHE} E$ ($\sigma_{SHE}$ is the SHC), the orbital current $J_O = \sigma_{OHE} E$ is transverse to the electrical current and satisfies the symmetry requirements for the X-parameter. The orbital Hall conductivity $\sigma_{OHE}$ (OHC) is always positive, i.e. independent of the $d$-electron band filling. We thus consider the spin interaction Hamiltonian to be $(\vec{s}, \vec{L}) \sim \sigma_{SHE} \sigma_{OHE} \vec{E}^2$, where $\vec{s}$ and $\vec{L}$ are the spin and orbital angular momenta dynamically generated via the SHE and OHE, respectively. Such an interaction modulates the dynamical electrical conductivity, $\sigma$, which consists of two parts: $\sigma = \sigma_0 + \gamma(\vec{s}, \vec{L})$. Here, $\sigma_0$ is the samples' electrical conductivity, and $\gamma(\vec{s}, \vec{L})$ is the spin-dependent THz-induced contribution with $\gamma$ a scaling factor introduced to match the units. Combining Maxwell's equations and the expression for the dynamical conductivity ($\sigma$) leads to a wave equation of the form (see Supplementary Information for details)

$$\Delta \vec{E} + \frac{\varepsilon}{\mu c^2} \frac{\partial^2 \vec{E}}{\partial t^2} = \frac{4\pi}{\mu c^2} \sigma_0 \frac{\partial \vec{E}}{\partial t} + \frac{12\pi}{\mu c^2} \gamma \sigma_{SHE} \sigma_{OHE} \vec{E}^2 \frac{\partial \vec{E}}{\partial t}. \quad (1)$$

Equation (1) comprises two terms on its right side. The first term describes the screening effect of the fundamental radiation, while the second term relates to THG, with an amplitude proportional to the SHC. This term is able to explain the observed 60-fold increase in the THG intensity in the Pd film as the temperature decreases down to 5 K (as shown in Fig. 3(C)). The 4 nm thick metal layer residual resistivity ratio (RRR = $\frac{\sigma_0(300K)}{\sigma_0(5K)}$) was measured to be approximately 1.7, suggesting that the THG amplitude scales with $\sigma_0^4$. In the high resistivity regime, which is likely relevant for the 4 nm thin films, both the intrinsic SHC and OHC increase proportional to $\sigma_0^2$ [24]. As a result, the 64-fold increase in the THG intensity is what we would expect based on the last term in equation (1).

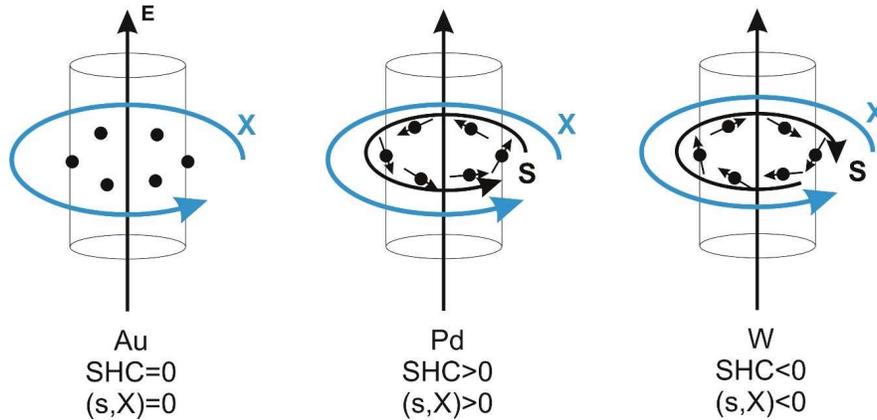

Fig. 4. Schematic representation of the THz electric field induced spin (S) and orbital (X) currents in Au (left), Pd (center) and W (right) samples. In all the metals, the electric currents are anticollinear to the applied electric field (E) and the orbital currents (X), generated via the orbital Hall effect, have the same

chirality. The spin currents (S), generated via the spin Hall effect, are zero for Au and have opposite chirality for the Pd and W samples.

**Conclusion**

Our experiments establish a THz nonlinear response occurring in the bulk of transition metal films, which is detectable through a THz third harmonic generation. THz THG processes can be observed in a wide range of metals, and we attribute them to the interactions between THz induced charge carriers, their spin and orbital degrees of freedom due to spin-orbital coupling. The THz THG amplitude was found to correlate with the spin Hall conductivity, while its phase is determined by the sign of the spin Hall angle. Our ab initio numerical DFT simulations of transition metals and estimates of the nonlinear carrier dynamics in Rashba-type media could not explain the observed behavior of the THz THG. We attribute the THz THG to the nonlinear interactions between THz electric field activated spin currents (via the spin Hall effect) and the THz field induced orbital currents (via the orbital Hall effect). These effects must be taken into account to understand the mechanisms of non-equilibrium and non-linear THz induced dynamics in transition metals on a microscopic level, and to understand the functionality of correlated materials at THz frequencies. Besides the fundamental aspects of nonlinear light-matter interactions in spin-orbit coupled systems, THz THG is of interest for practical applications, e.g. for the lithographic- and contact-free determination of ultrafast charge-spin interconversion, which forms the basis of emerging THz spintronics and THz orbitronics technologies.

**Methods**

**Sample fabrication**

The transition metal films were fabricated using dc magnetron sputter deposition at $3\times10^{-3}$ mbar Ar atmosphere in an ultrahigh-vacuum ($4\times10^{-9}$ mbar) BESTEC system. We used 1-mm-thick double-side polished quartz ($SiO_2$) substrates. A 10-nm-thick $SiO_x$ layer, grown using rf sputter deposition, served as a cap layer to prevent the surface oxidation of the metallic films. All metallic layers were deposited at room temperature, with the use of a rotating sample holder to ensure the uniform growth of the metallic layers.

$Ni_{80}Fe_{20}$ and Fe films with a thickness of 4 nm were grown on single-crystalline MgO(001) substrates by molecular beam epitaxy (MBE) in a growth chamber with a base pressure below $3\times10^{-10}$ mbar at ambient temperature. Both films were capped by a 3-nm-thin Au layer. For the $Ni_{80}Fe_{20}$ we used electron beam evaporation, while Fe and Au were evaporated from an effusion cell with a flux of high stability. Before the deposition, the MgO substrates were preheated at 100°C at a pressure of $4\times10^{-7}$ mbar for 2 hours in the load lock chamber of the MBE system.

The thicknesses of all layers were controlled via the deposition time. Before the sample fabrication, the growth rate of each individual material was calibrated using X-ray reflectivity characterization of the corresponding film.

Thin films of *25-nm-thick* Bi were grown on $SiO_2$ quartz glass substrates (CrysTec GmbH, Germany) and capped with a 2-nm-thick Au layer to prevent oxidation. The deposition of Bi/Au bilayers was done using RF magnetron sputtering at room temperature (base pressure: better than $10^{-7}$ mbar; Ar sputter pressure: $10^{-3}$ mbar; deposition rate: 0.3 nm/s).

**Experimental setup**

Time-domain spectroscopy was conducted utilizing both laser-based and accelerator-based terahertz sources. The laser-based THz source was based on a tilted pulse front generation (TPFG) scheme using a Ti:Sapphire laser system with 9 mJ pulse energy, 35 fs pulse duration at 800 nm central wavelength, operated at 1 kHz repetition rate. The THz pulses from TPFG scheme are initially broadband with frequency content spanning from 0.2 THz to 1.5 THz. Bandpass filters were employed for the conversion of broadband terahertz radiation into a narrowband spectrum. For temperature dependence of THz THG, we used the accelerator-based THz source TELBE located at Helmholtz-Zentrum Dresden-Rossendorf.

**Acknowledgments:** The authors are grateful to Thomas Naumann and Andreas Henschke for technical support. Parts of this research were carried out at ELBE at the Helmholtz-Zentrum Dresden-Rossendorf e.V., a member of the Helmholtz Association.

**Funding:** M.L. and P.W. acknowledge support from ERC Consolidator Grant No. 724103. D.M. and P.M. acknowledge support from the German Research Foundation (DFG) Grants No. MA5144/22-1, MA5144/24-1, MA5144/33-1 and European Commission (project REGO; ID: 101070066). R.S. and A.L. acknowledge support from the German Research Foundation (DFG) Grants No. LI3827/2-1 (464974971).

**Author contributions:** S.K., and R.S. initiated the project. S.K., R.S., M.L., P.W., D.M., J.L., and C.O. designed and planned the work, R.S., P.M., D.M., and A.L. contributed to sample preparation. S.K., R.S., I.I., T.V.A.G.O., A.P., A.A., G.L.P., and J.-C.D. carried out the experiment. M.L., P.W. and M.S. performed the DFT calculations. C.O. performed the estimations of surface states contribution to the THz THG. S.K., P.W., D.M., T.C., J.F., J.L., and A.L. acquire the funding for this work. S.K., R.S., M.L., P.W., and C.O. wrote the manuscript with input from all authors. All authors provided critical feedback and helped shape the research, analysis and manuscript.

**Competing interests:** Authors declare that they have no competing interests.

**Data and materials availability:** All data are available in the main text or the supplementary information.